\documentclass[aps,prd,preprint,nofootinbib]{revtex4-1}
\usepackage{amsmath,amssymb,graphics,graphicx,color}
\usepackage{scrextend}
\usepackage{hyperref}
\usepackage{float}
\usepackage{xcolor}
\usepackage[utf8]{inputenc}
\usepackage[english]{babel}

\begin{document}
\preprint{NCTS-HEPAP/2102}
	
\title{Non-standard Neutrino and $Z'$ Interactions  at the FASER$\nu$ and the LHC}
\author{ Kingman Cheung$^{a,b,d}$,  C.J. Ouseph$^{a,b}$, TseChun Wang$^c$,}
\affiliation{
	$^a$ Department of Physics, National Tsing Hua University, Hsinchu 300, Taiwan\\
	$^b$ Center for Theory and Computation, National Tsing Hua University, Hsinchu 300, Taiwan \\
	$^c$ Physics Division, National Center for Theoretical Sciences, Taipei 10617, Taiwan \\
	$^d$ Division of Quantum Phases and Devices, School of Physics, Konkuk University, Seoul 143-701, Republic of Korea
}
\date{\today}
\begin{abstract}
We study the impact of non-standard neutrino interactions in the context of a new gauge boson $Z'$ in neutral-current deep-inelastic scattering performed 
in ForwArd Search ExpeRiment-$\nu$ (FASER$\nu$) and in monojet production at the 
Large Hadron Collider (LHC).
 We simulate the neutral-current deep-inelastic neutrino-nucleon scattering 
 $\nu N \rightarrow \nu N$ at FASER$\nu$ in the presence of an additional $Z'$ boson,
 and estimate the anticipated sensitivities to the gauge coupling in a wide 
 range of $Z'$ mass.  
 At the LHC, we study the effect of $Z'$ on monojet production,
 which can be enhanced in regions with large missing transverse momenta. 
 We then use the recent results from ATLAS with an integrated luminosity of 139 fb$^{-1}$ to improve the limits on the gauge coupling of $Z'$. 
 We interpret such limits on $Z'$ gauge couplings as bounds on 
 effective non-standard neutrino interactions.
We show that the FASER$\rm \nu$ and the LHC results cover the medium and high energy scales, respectively, and complement one another.
\end{abstract}
\maketitle
\section{Introduction}

Determining the properties of neutrinos is important in the context of searching for physics beyond the standard model and understanding the universe. There are several motivations of these measurements. First, the neutrino oscillation indicates that the neutrinos are light but not massless, in contrast to the prediction 
by the standard model (SM). This rock-solid fact motivates physicists to search for any other neutrino properties, which are beyond the standard model (BSM). Second, the precision of parameters in the neutrino physics needs improvements. This is because neutrinos are hard to be detected, compared to the other charged fermions. In these measurements, statistical uncertainties are usually dominating over other sources of uncertainties. 
Therefore, even if there are truly some BSM features, our precision might be not good enough to catch these features. And finally, neutrino is one of the most abundant particles in the universe since the Big Bang nucleosynthesis (BBN). As a result, the property of neutrinos is an essential factor in the evolution of the universe, \textit{e.g.}~the neutrino-neutrino self interaction might affect the measurement of Hubble constant $H_0$
\cite{Blinov:2019gcj,Cyr-Racine:2013jua,Kreisch:2019yzn,Ghosh:2019tab}.

To catch the BSM features in neutrino physics, physicists have considered a variety of experimental configurations. One approach,
which has the least uncertainty, is to observe the neutrinos in collider experiments. 
Several proposals have been discussed, \textit{such as}: Search for Hidden Particle (SHiP)
\cite{DeLellis:2015usa,SHiP:2015vad} and 
ForwArd Search ExpeRiment (FASER)~\cite{FASER:2019dxq,FASER:2019aik}, etc.
Both SHiP and FASER are proposed to detect neutrinos and 
long-live particles from CERN. In addition to the FASER main detector, the 
sub-detector in the front is a $1.2$-ton 
tungsten detector -- FASER$\nu$~\cite{FASER:2019dxq}. 
With high neutrino luminosity ($2\times 10^{11}\nu_e$, $6\times10^{12}\nu_\mu$, $4\times 10^9\nu_\tau$  during LHC-Run3), FASER$\nu$ provides an optimal window for precision measurements of neutrino properties of all flavors at 
the medium-high energy scale ($600$~GeV to $1$~TeV). This can also be used to search for BSM physics, \textit{e.g.}~new interactions. The main detector in FASER is now taking data. When the LHC restarts in 2022. FASER$\nu$ is expected to measure the neutrinos during the next period of LHC operation from 2022 to 2024.

The most distinct feature of the FASER$\nu$ experiment is the unique energy range of neutrinos that it can cover. The ICECUBE focuses on very high-energy neutrinos with energy 10 TeV to 1 PeV, and the LHC covers from hundreds of GeV to a few TeV. On the other hand, the short- and long-baseline experiments cover mostly around MeV up to a few GeV. There are no precise measurements of neutrino scattering in a few tens of GeV to a few hundreds of GeV region. FASER$\nu$ based on the neutrino flux coming off the LHC opens such a unique window in this energy range.

One of the simplest extensions to the SM is to add an extra $U(1)$ gauge symmetry,
which results in a new neutral gauge boson $Z'$. Such a $Z'$ boson can couple to
the SM fermions or simply hidden, depending on the construction.
It is also motivated by some theoretical models (\textit{e.g.}~dark matter
models~\cite{He:1991qd,Shepherd:2009sa}). 
However, without any hint about the energy scale of this new physics, we can only treat the mass of $Z'$ ($M_{Z'}$) as a free parameters. A systematical search in a wide energy range is
therefore very important. As this $Z'$ might be a generator of a new symmetry of flavor \textit{such as} the 
$\mu$-$\tau$ symmetry, the neutrino detection with all flavors is an advantage to test the $Z'$ models of this kind. 
We see that FASER$\nu$, with the capability of distinguishing the flavors of neutrinos,
will play an important role in the $Z'$ search and test for the flavor structure
in the $Z'$ interactions.
We will further investigate these features in this work.

On the other hand, the LHC monojet production can cover effectively the mass range
from a few hundred GeV up to a few TeV.
The LHC monojet data can put stringent constraints on the $Z'$ gauge coupling. We use the most updated monojet data with 139 fb$^{-1}$ luminosity \cite{ATLAS:2021kxv},
and obtain the best limit on the gauge couplings $g_q g_{\nu}$, which can be translated to
the effective 
$\epsilon (\bar q \gamma^\mu q ) ( \bar \nu_L \gamma_\mu \nu_L )$.
Considerable improvement over previous works is demonstrated here. 
Nevertheless, the monojet data is not sensitive to the flavor of the neutrinos, and therefore the $\epsilon$ is the sum of contributions from all three flavors, in contrast to low-energy oscillation experiments.

The organization of the work is as follows. In the next section, we briefly introduce the theoretical aspects of $Z'$, 
the relevant phenomenology, and the current status. In Sec. III, we show the effects of $Z'$ interactions on LHC monojet 
production and obtain the limits on the effective NSI. In Sec. IV, we study the sensitivities of $Z'$ interactions achieved at FASER$\nu$. 
In Sec. V, we show the complementarity of LHC monojet production and FASER$\nu$ in the coverage of mass range of $Z'$.
Finally, we give our conclusions in Sec. VI.

\section{The $Z'$ model and Non-standard Neutrino Interactions}


%
Renormalizable interactions of the $Z'$ with flavor-conserving quark and neutrino interactions can be written as 
\begin{equation}
\label{zint}
     {\cal L}_{Z'} = - \left( g_{\nu} \bar{\nu} \gamma^{\mu} P_L \nu \;+\;
                            g_{q}\bar{q}\gamma^\mu q \right) \; Z'_{\mu} \;.
\end{equation}
Here we assume that $q=u, d$ have equal coupling strength $g_q$ and $\nu = \nu_e. \nu_\mu, \nu_\tau$ have equal strength $g_\nu$. 
In this simplified $Z'$ model, we assume that the coupling strengths to the
left- and right-handed $u,d$ quarks are the same, and so are 
the coupling strengths to the three flavors of neutrinos,
as production of high-energy neutrinos is not sensitive to the flavors
of neutrinos. Nevertheless, the results can be easily extended to 
non-universal coupling strengths.

Although we use a simplified  $Z'$ model in our working procedures,
there are still a number of existing constraints on general $Z'$ models.
We briefly discuss in the following.
\begin{enumerate}
    \item The Big Bang Nucleosynthesis (BBN) places constraints on the mass of the boson 
    $M_{Z'} \lesssim 5$ MeV~\cite{Huang:2017egl}.
    
    \item Supernova cooling also leads to substantial effects in the observed supernova neutrino spectrum, 
    which implied the $Z'$ coupling to be as small as $g_\nu \sim 10^{-10}$ for 
    $M_{Z'} \alt 30\; {\rm MeV}$~\cite{Dent:2012mx,Harnik:2012ni}.
    
    \item The branching ratio for $K^0_L\rightarrow \pi^0 Z'$ leads to a bound $g_q\lesssim 10^{-8}$ 
    for $M_{Z'} = 100-200\, {\rm MeV} $~\cite{Nelson:1989fx}.
    
    \item  The measurement of $\eta\rightarrow\pi^0\gamma\gamma$ 
      gives a bound $g_q\lesssim 10^{-5} - 0.01$ for $M_{Z'}$ ranging from 200 to 600 MeV
      \cite{Tulin:2014tya}. 
    Other measurements on the branching ratios of $\eta'\rightarrow\pi^0\pi^+\pi^-\gamma$,
    $\psi\rightarrow K^+,K^-$, and $\Upsilon\rightarrow$ hardons give a bound 
    $g_{q}\lesssim 0.01-0.1$ for $M_{Z'}=0.5,5.5$ and 9.8 GeV.
    
    \item BaBar put a constraint on the coupling strength of electron to $Z'$ $g_e \lesssim 3.3\times10^{-2}$ from the process $e^+ e^- \rightarrow\gamma Z'$ with $M_{Z'} \lesssim 10\, {\rm GeV}$~\cite{BaBar:2008aby,Essig:2013vha}.
    
    \item Borexino placed a bound $g_{e,\mu}\lesssim \mathcal{O}(10^{-2})$ 
      for $M_{Z'} \sim 1\,{\rm GeV}$.
       Furthermore, for a very light $Z'$ of mass $M_{Z'}\sim 1\,{\rm MeV}$
       the constraint becomes more stringent
       $g_{e,\mu}\lesssim\mathcal{O}(10^{-5})$~\cite{Harnik:2012ni}.
\end{enumerate}

\subsection{Non-standard Neutrino Interactions}

It is clear from the above discussion that the constraints on tree-level 
couplings of $Z'$ to neutrinos, charged leptons, and quarks are quite 
stringent for $M_{Z'}\lesssim 1 {\rm GeV}$. 
In the following sections, we investigate the effects of the 
non-standard neutrino interactions (NSI) or $Z'$ 
interactions on monojet production at the LHC and NC scattering at FASER$\nu$,
which signify a very high energy and a medium energy scale, respectively.

The pursuit of NSI's is one of the main goals of current and future neutrino experiments. The NSI's can be categorized into charged current and neutral current ones.
Specifically, we are looking at neutral current NSI's\cite{Proceedings:2019qno}

\begin{equation}\label{eq1}
  {\cal L}_{NC} = -2\sqrt{2}G_F \, \sum_{f,P,\alpha, \beta} \, \epsilon_{\alpha\beta}^{f,P}\,
   \left ( \bar \nu_\alpha \gamma^\mu P_L \nu_{\beta} \right ) \,
    \left ( \bar f \gamma_\mu P  f \right)  \; ,
\end{equation}
where $G_F$ is the Fermi constant,
$\alpha,\beta$ are flavor indices, $(f,f') =(d,u)$,  $P = P_L$ or $P_R$ is the chirality projection
operator. The parameters $\epsilon_{\alpha\beta}^{f,P}$ quantify 
the strength of the NC NSI's, 
$\alpha, \beta=e,\mu,\tau$, $f=u,d$. For simplicity we only consider the 
flavor-conserving interactions on the quark leg, while the neutrino leg
allows for changes in neutrino flavors.
Note that the neutrino field $\nu_L$ originates from the lepton doublet $L$, such 
that the above interactions can be generated from SM gauge invariant higher dimensional 
operators, such as 
\begin{equation}\label{eq2}
  - \frac{1}{\Lambda^2}\, \left(\bar L_\alpha \gamma_\mu L_\beta \right )\, 
   \left [ \bar Q \gamma^\mu P_L Q  +
     \bar u_R \gamma^\mu P_R u_R  + \bar d_R \gamma^\mu P_R d_R \right ]
\end{equation}
where $L$ is the lepton doublet, $Q$ is the quark doublet, $u_R,d_R$ are the 
quark singlets. We can then equate to obtain
\begin{equation}
\epsilon^{f,P}_{\alpha\beta} =  \frac{1}{2\sqrt{2}G_F \Lambda^2} \;.
\end{equation}
Since in this work we deal with the effects of the NC NSI's at the FASER$\nu$ and the LHC, 
one may concern about the validity of the effective operators in Eq.~(\ref{eq1}). The simplified $Z'$ model in Eq.~(\ref{zint}) converges back to Eq.~(\ref{eq1}) when $m_{Z'} \to \infty$.

Similarly, when the square of momentum transfer $\hat s, |\hat t|$ are much smaller than $M_{Z'}$, the ratio $\epsilon^{f,P}_{\alpha\beta}$ can be approximated by $\epsilon_{eff}$,
\begin{equation}
\label{eps}
   \epsilon_{eff}   = \frac{g_q g_\nu}{2\sqrt{2}G_F M^2_{Z'}} \;.
\end{equation}
Straightly speaking here the $\epsilon_{eff}$ is not the same as 
$\epsilon^{f,P}_{\alpha\beta}$ of Eq.~(\ref{eq1}), but for ease of 
comparison to those limits obtained at low energies.  
In this work, we first work out
the sensitivity constraints in terms of coupling strengths of $Z'$, and
then later translate back to the effective coupling $\epsilon_{eff}$'s.

\section{Effects of $Z'$ on Monojet Production}

A number of new physics models, such as large extra dimensions, invisibly decaying scalar bosons,
sterile neutrinos and dark matter models,
can give rise to missing-energy signals at the LHC, other than the
active neutrinos.  The visible object in such events would be the single jet radiating off
the initial quark legs, giving rise to monojet events plus large missing energy. 
In the current $Z'$ model, the $Z'$ boson can be produced associated with a jet, 
followed by the $Z'$ decay into neutrinos. Thus, the signature is a single jet plus large missing
energy.  In the following, we calculate the production rates of monojet production due to the 
$Z'$ interactions. Without loss of generality we assume the $Z'$ boson couplings to $u$ and $d$ 
are the same, and do not couple to other generations. 
We can easily extend to different $Z'$ couplings in expense of more independent parameters. 
After computing the production rates for monojet events, we can then 
use a recent experimental result on monojet production \cite{ATLAS:2021kxv} 
to put bounds on the product of couplings $(g_q g_\nu)$. 
Note that production of monojet events has been studied to test effective neutrino-quark 
interactions \cite{Friedland:2011za,Pandey:2019apj,Babu:2020nna,Liu:2020emq}.
An improvement on the constraints can be achieved from
previous works because we have used the most recent result on monojet production \cite{ATLAS:2021kxv}.

\begin{figure}[th!]
	\centering
	\includegraphics[width=17cm,height=9cm]{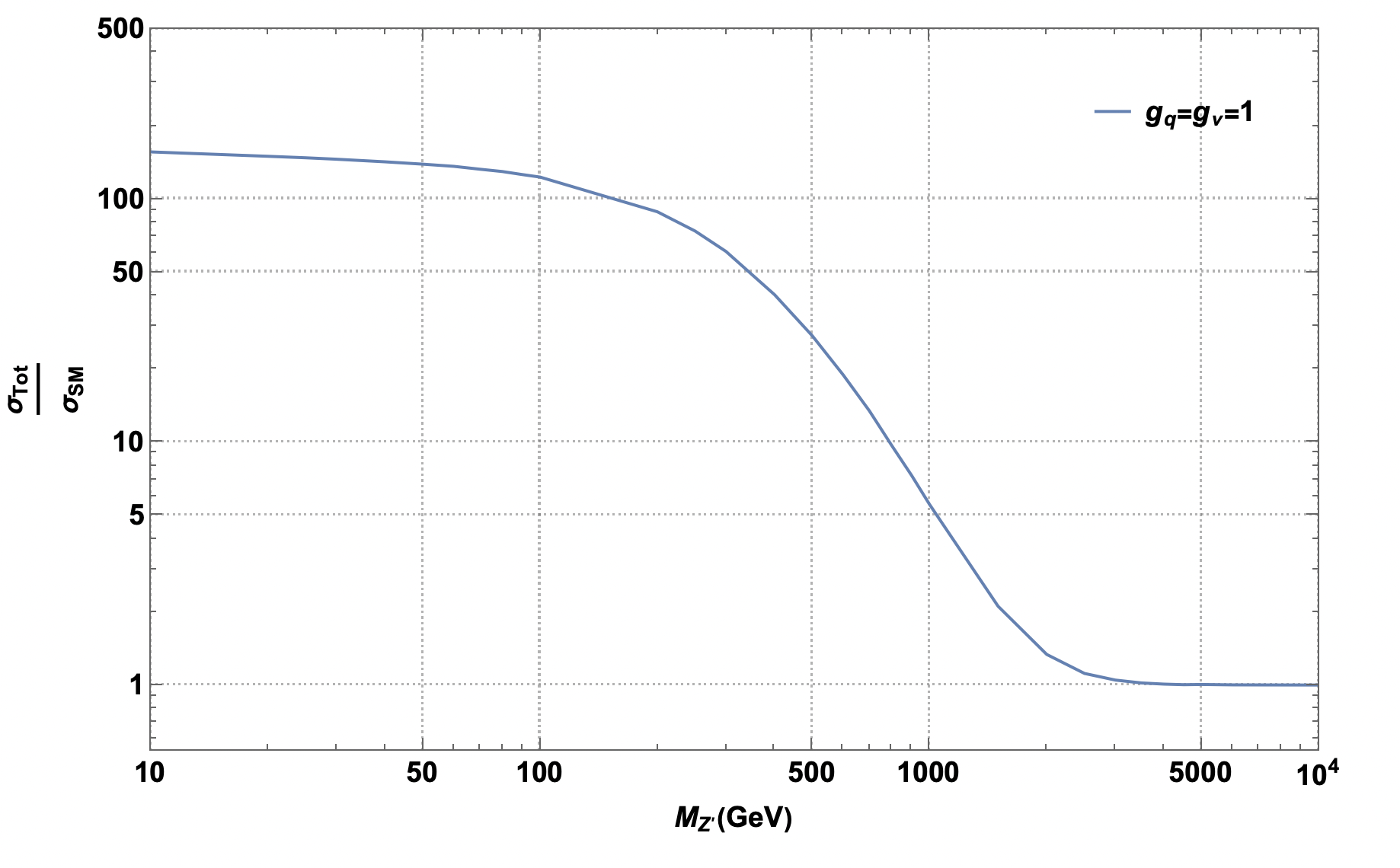}
	\caption{ \small \label{fig1}
	The ratio of the cross sections $\sigma_{\rm Tot}/\sigma_{\rm SM}$ versus the mass of the 
	$Z'$ boson, in which we have used $g_q = g_\nu = 1$.}
\end{figure}

The process that we calculate is 
\[
  p p \to \bar \nu_\beta \nu_\beta + j ,
\]
where we sum over all three neutrino flavors assuming their couplings $g_\nu$ to be the
same, and $j$ refers to either $q,\bar q, g$.  The contributing Feynman diagrams include
the SM $Z$ boson and the $Z'$ boson exchanged in $s$-channel.  
When the mass $M_{Z'} \to\infty$ the SM result is restored. 
In principle, the $Z$ and $Z'$ diagrams interfere with each other, such that the 
interference term is proportional to the couplings $(g_q g_\nu)$ while the sole $Z'$ 
contribution is proportional to $(g_q g_\nu)^2$. 

In the calculation, we generate the aforementioned process using
MadGraph5aMC@NLO~\cite{Alwall:2011uj,Alwall:2014hca}
with the model file generated by the effective Lagrangian in Eq.~(\ref{zint}), 
followed by parton showering and hadronization with PYTHIA8~\cite{Sjostrand:2014zea},
detector simulations  carried out by Delphes3 package~\cite{deFavereau:2013fsa}. 
The total cross-section $\sigma_{\rm Tot}$ for $p p \to  \nu\bar {\nu}+1j $ can be 
expressed as follows
\begin{equation}
\sigma_{\rm Tot}=\sigma_{Z'} + \sigma_{\rm Int} + \sigma_{\rm SM} \;,
\end{equation}
where $\sigma_{Z'}$ is the cross-section of the aforementioned process only with 
the $Z'$ propagator, $\sigma_{\rm Int}$ is the interference term and 
$\sigma_{\rm SM}$ is the 
standard model cross-section.  We show the ratio of $\sigma_{\rm Tot} / \sigma_{\rm SM}$ 
in Fig.~\ref{fig1}. It is clear that the $\sigma_{\rm Tot}$ approaches $\sigma_{\rm SM}$
as $M_{Z'}$ becomes very large.
Note that the total decay width of the $Z'$ boson is assumed to be $\frac{\Gamma_{Z'}}{M_{Z'}}=0.1$.

\subsection{Sensitivity reach on parameter space of the $Z’$ model and NSI's}
Here we derive the bounds on the product of the $Z'$ couplings $(g_q g_\nu)$ as a function of 
$M_{Z'}$ based on a recent result on monojet production by the ATLAS experiment \cite{ATLAS:2021kxv}. 
Later, our goal is to translate such constraints into the conventional NSI parameters $\epsilon_{eff} = \epsilon_u=\epsilon_d$ defined in Eq.~(\ref{eq1}). 

We follow closely the experimental cuts outlined in the ATLAS paper \cite{ATLAS:2021kxv} in order to 
directly use their upper limits on the monojet production cross sections. 
Their results were based on the monojet search at 13 TeV with an integrated luminosity of 
139 $\rm fb^{-1}$~\cite{ATLAS:2021kxv}. Events are selected with $E_T^{\rm miss} > 200\,{\rm GeV}$, 
a leading jet with $p_T>150\,{\rm  GeV}$ and $|\eta|<2.4$ and upto three jet with 
$p_T>30 \, {\rm GeV}$ and $|\eta|<2.8$, as well as  
additional cuts specified in \cite{ATLAS:2021kxv}. Jets are defined with the anti-$k_t$  jet algorithm 
with a cone size $R = 0.4$. 

\begin{figure}[th!]
	\centering
	\includegraphics[width=17cm,height=7cm]{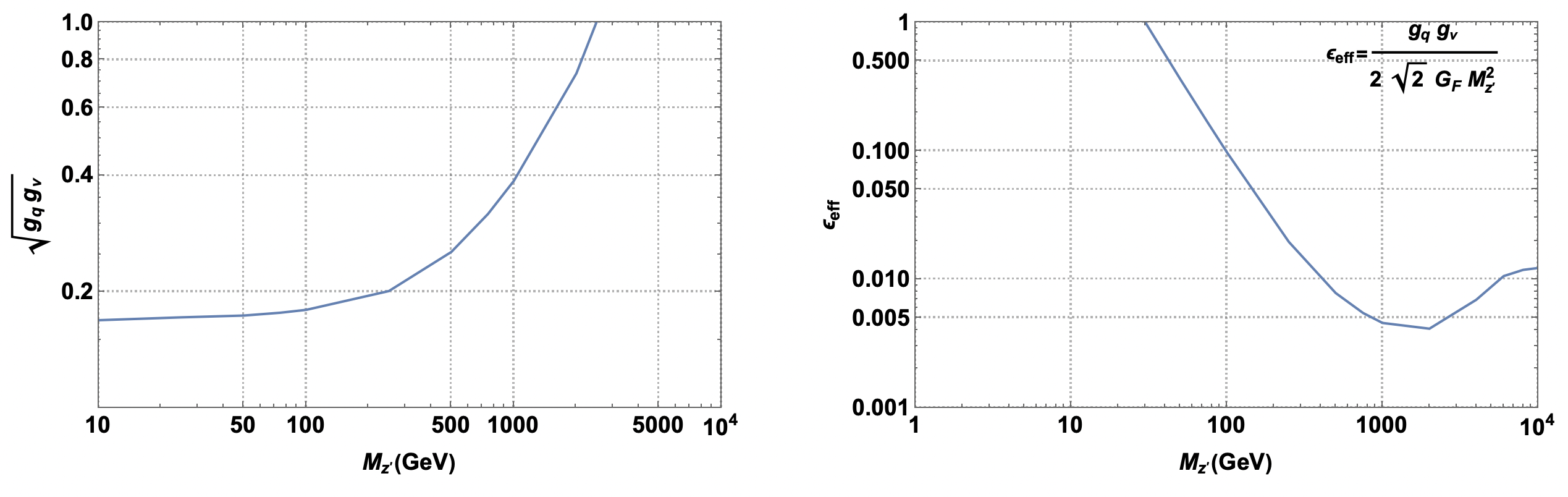}
	\caption{
	 \small \label{fig2}
	Left: Monojet bounds on the product of couplings $\sqrt{g_q g_\nu}$ versus the $Z'$ mass. 
	Right: Constraints on $\epsilon_{eff}$ of the NSI's translated from $(g_q g_\nu)$ using  Eq.~(\ref{eps}) 
	versus $M_{Z'}$. Here we have assumed $\epsilon_{ee} = \epsilon_{\mu \mu} = \epsilon_{\tau\tau} = \epsilon_{eff}$. 
	}
\end{figure}

With all the acceptance cuts the same as Ref.~\cite{ATLAS:2021kxv}, we still need the overall
efficiency in order to obtain the event rates to compare with the experimental results.
We rely on an information given in Ref.~\cite{ATLAS:2021kxv}. A signal model with 
an axial-vector gauge boson $Z_A$, via which a pair of 
dark matter particles $\chi$ can be produced in $s$-channel, was investigated. 
The reported "acceptance $\times$ efficiency'' in the kinematic region 
EM0 \footnote{Here $p_T^{\rm recoil}$ is the same as $E_T^{\rm miss}$ for signal models.}
($p_T^{\rm recoil} =200 - 250$ GeV) 
was 13\%.  Since the event  
topology of such a signal ($pp \to Z_A +j \to \bar \chi \chi +j$) 
is similar to our signal ($pp \to Z' + j \to \bar \nu \nu +j$), 
we can then compare our acceptance to their value of ``acceptance $\times$ efficiency''.
Therefore, we obtain an efficiency of $0.582$, which is then applied to
all our event rates.  We have calculated bounds of $\sqrt{g_q g_\nu}$ 
using the 95\% C.L. upper limits on the signal event rates in 
a number of kinematic regions defined in Ref.~\cite{ATLAS:2021kxv} (see Table 9 of Ref.~\cite{ATLAS:2021kxv}).
The resulting limits are within a factor of two among one another. 
We show in the left panel of Fig.~\ref{fig2} the bounds on  $\sqrt{g_q g_\nu}$ based on the 
95\% C.L. upper limit on the observed event rate $S^{95}_{\rm obs} = 11937$
in the kinematic region IM3 ($p_T^{\rm recoil} > 350 $ GeV) \cite{ATLAS:2021kxv}.

Next we can translate the bounds on $\sqrt{g_q g_\nu}$ to $\epsilon_{eff}$ using Eq.~(\ref{eps}). 
The bounds on $\epsilon_{eff}$ are shown on the right panel of Fig.~\ref{fig2}. 
We can see that the best limit on $\epsilon_{eff}$ appears around $M_{Z'} \sim 2 $ TeV.
The bound becomes less stringent as the $Z'$ mass increases, because the $Z'$ becomes 
more difficult to be produced directly. 

\section{Effects of $Z'$ and Neutral-current NSI's interactions at FASER$\nu$}

FASER \cite{FASER:2019dxq,FASER:2019aik}
is an approved experiment located about 480 m away from the interaction 
point (IP) of the ATLAS detector down along the direction of the proton beam. It is well known 
that huge number of hadrons, such as pions, kaons and other hadrons, are produced along the 
beam direction. These hadrons will decay during the flight, thus producing a lot of neutrinos of
all three flavors at very high energy up to a few TeV.   

There is a proposed new component, called FASER$\nu$ \cite{FASER:2019dxq}, to be put in front of the FASER detector.
It is an $\rm 25 cm\times25cm\times1.5m$ emulsion detector, consisting of 1000 layers of emulsion films 
interleaved with 1-mm-thick tungsten plates with mass 1.2 tons.  The main goal of FASER$\nu$ is to 
distinguish various flavors of neutrinos. Indeed, it can measure the flux of electron, muon, and tau neutrinos
coming off from the IP of the ATLAS detector, which can be done by detecting the charged lepton coming off
the charged-current (CC) scattering. Notably, muon nutrino is the most abundant due to production of charged
pions and kaons while tau neutrino is the least as it requires at least the heavier mesons. like $D_s$ meson,
in order to decay into $\tau \nu_\tau$. 
On the other hand, it is also feasible to measure the neutral-current (NC) scattering of the neutrinos 
\cite{FASER:2019dxq} making use of the emulsion detector, although the detection of NC interactions is 
somewhat more difficult than the CC one. 
The total cross-section ($\sigma_{\nu N}$) of the NC scattering at the FASER$\nu$ detector can be 
expressed as $\sigma_{\nu N} = n \sigma_{\nu n} + p\sigma_{\nu p}$, where $\sigma_{\nu n}$ and 
$\sigma_{\nu p}$ are the neutrino-neutron and neutrino-proton scattering cross sections, and $n$ and $p$ 
are the number of neutrons and protons in the tungsten atom, respectively.

In this section we compute the sensitivity of FASER$\nu$ to the NC NSI's due to physics beyond the SM.
Similar to the last section, we use the same simplifed $Z'$ model (see Eq.~(\ref{zint}) ) to calculate the
sensitivity reach at FASER$\nu$.  The effect of $Z'$ is similar to that at the LHC, other than the fact that
the $Z$ and $Z'$ bosons are exchanged in $t$-channel in the NC deep-inelastic scattering,
such that the most significant effect of $Z'$  appears in the small $M_{Z'}$ region. 
The SM result is restored as $M_{Z'} \to \infty$. 
We estimate the 95\% C.L. sensitivity reach on the parameter space of the $Z'$ model. 
We show that the best sensitivity can be achieved in the small $M_{Z'}$ region that it is 
highly complementary to that obtained by monojet production at the LHC.

\subsection{$Z'$ Interactions at FASER$\nu$}

\begin{figure}[ht!]
	\centering
	\includegraphics[width=17cm,height=10cm]{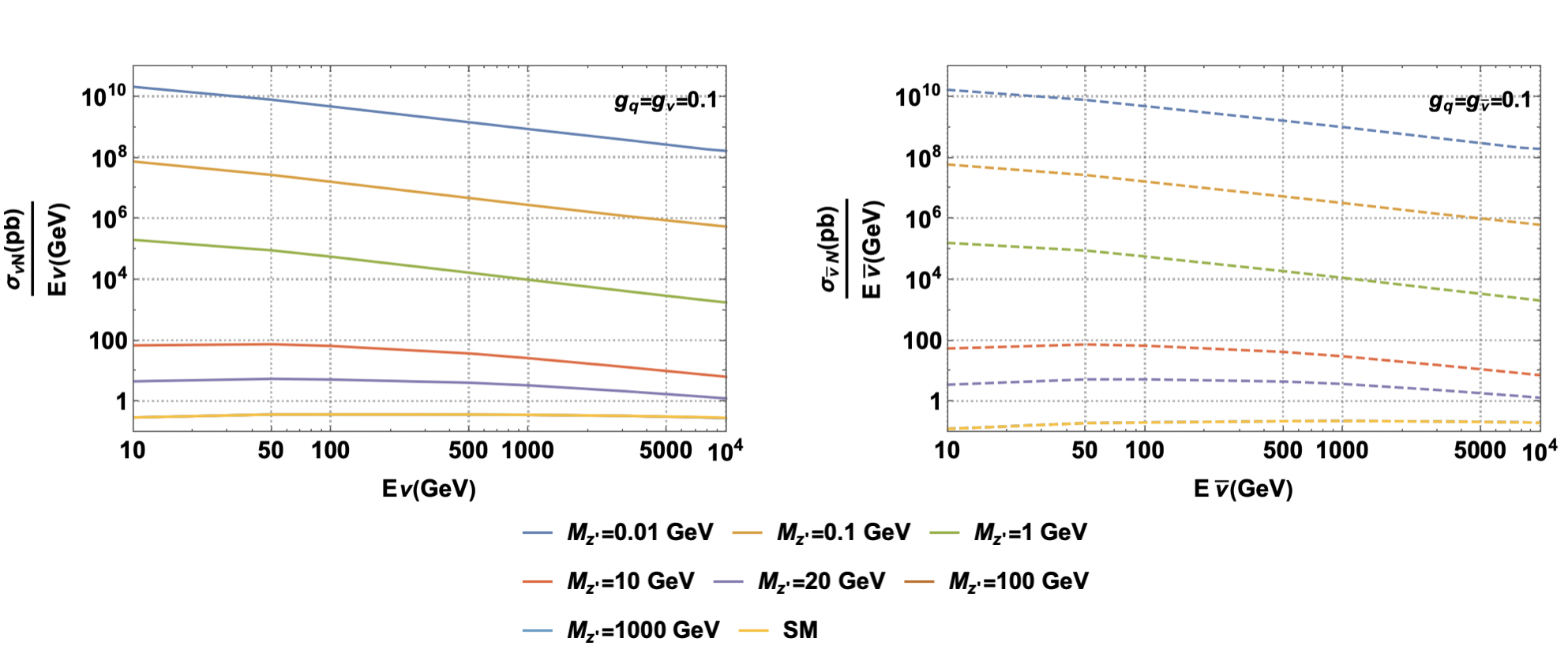}
	\caption{ \small \label{fig3}
	Left: Deep-inelastic neutral-current scattering cross section normalized by the energy $E_\nu$ 
	of the incoming neutrino beam. Right: the same as the left but with anti-neutrino beam. 
	Here N is the tungsten nucleus. We have used the CTEQ6L1\cite{Pumplin:2002vw}  for 
	parton distribution functions.  We have set $g_q = g_\nu = 0.1$. 
	}
\end{figure}

The square of the Feynman amplitude for the subprocess $\nu (p_1) q (p_2) \to \nu (k_1) q(k_2)$, where 
$q=u,d$ and the 4-momenta of each particle is shown in parenthesis, is given by
\begin{equation}
 \sum \left | {\cal M} \right |^2 = 4 \hat u^2 | M^{\nu q}_{LL} |^2  + 4 \hat s^2 |M^{\nu q}_{LR} |^2
 \end{equation}
where the reduced amplitudes $M^{\nu q}_{L\beta}$ are given by
\begin{equation}
     M^{\nu q}_{L \beta} \left ( \nu (p_1) q (p_2) \to \nu (k_1) q(k_2) \right ) = 
    \frac{e^2 g_Z^{\nu}  g_Z^{q_\beta} } { \sin^2\theta_w \cos^2 \theta_w} \, \frac{1}{\hat t - M_Z^2 }
    + \frac{g_\nu g_{q_\beta} }{ \hat t - M_{Z'}^2 } \;,
\end{equation}
where $\beta=L,R$, $g_Z^{f_L} = T_{3f} - Q_f \sin^2 \theta_w$, and 
$g_Z^{f_R} = - Q_f \sin^2 \theta_w$ are the SM $Z$ couplings to 
the fermion $f_L$ and $f_R$, and $\theta_w$ is the Weinberg angle. 
Here $\hat s, \hat t, \hat u$ are 
the usual Mandelstam variables. In our $Z'$ model, the couplings
$g_{q_L}$ and $g_{q_R}$ are the same. 
It is easy to see that when $M_{Z'} \to \infty$ the SM result is restored.
We used MadGraph5aMC@NLO~\cite{Alwall:2011uj,Alwall:2014hca} for fixed target 
deep-inelastic neutrino-nucleon 
scattering computation. We build the model file for Eq~(\ref{zint}) using
Feynrules~\cite{Alloul:2013bka}.
We consider the $Z'$ mass ranging from 0.01 GeV to 10 TeV and show the scattering cross section 
normalized by $E_\nu$ in Fig.~\ref{fig3}.
The process cross-section $\sigma_{\nu {\rm N}}$ decreases with $M_{Z'}$. 
At heavy $M_{Z'}$ mass regime the total cross section approaches to 
the standard model values, 
which was already reported in \cite{FASER:2019dxq,IceCube:2017roe,Ismail:2020yqc,Cooper-Sarkar:2011jtt}. We used the values of neutrino flux
\cite{FASER:2019dxq} for 
the evaluation of neutrino-nucleus interaction with FASER$\nu$

\begin{figure}[hbt!]
	\centering
	\includegraphics[width=17cm,height=13cm]{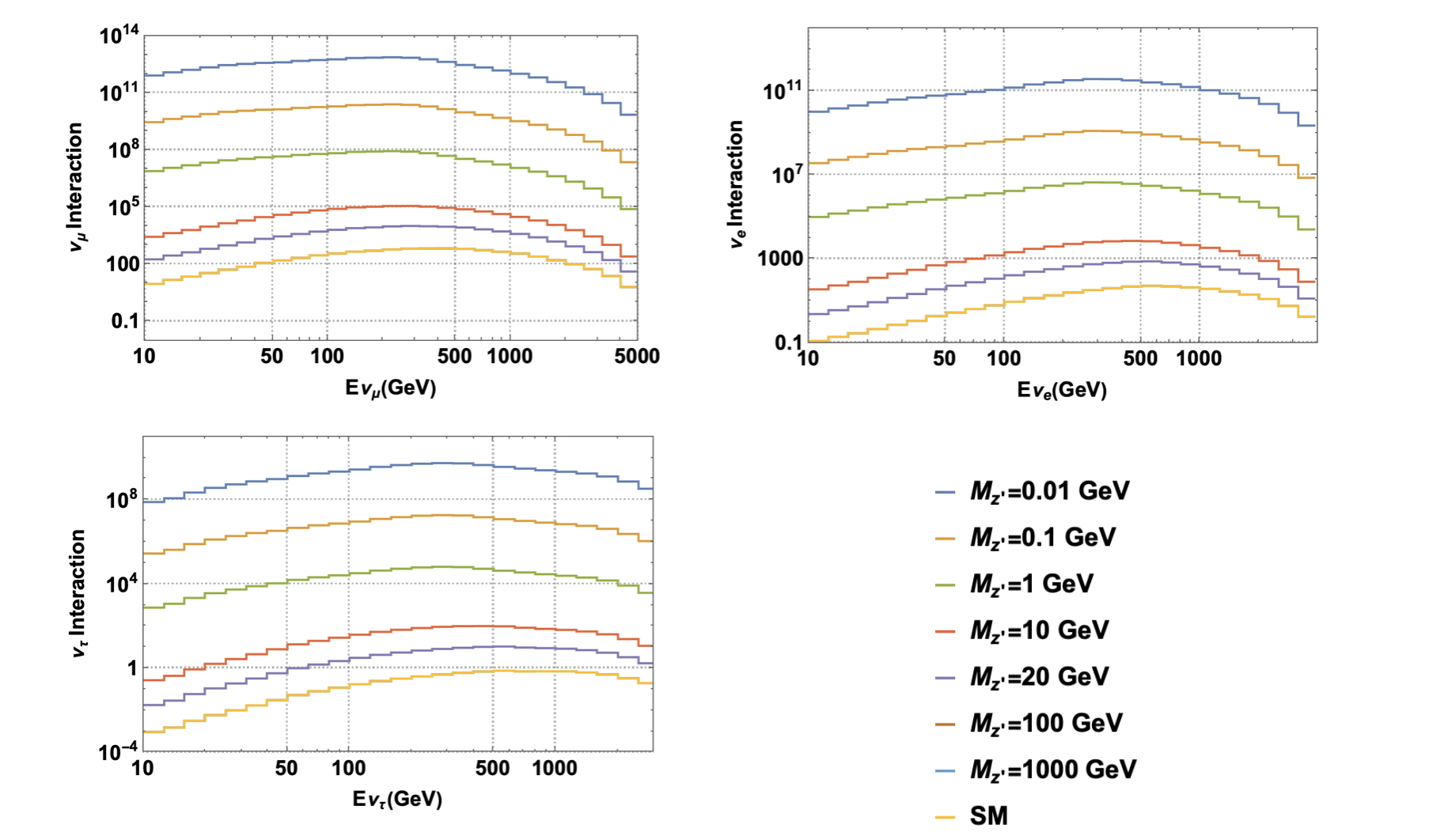}
	\caption{ \small \label{fig4}
	The energy spectrum of neutrinos with NC interactions mediated with $Z$ and $Z'$ in a 
	1-ton tungsten detector with dimensions $\rm 25\, cm\times25\, cm\times1\, m$ 
	centered on the beam collision axis at the FASER location at the 14 TeV LHC with 
	150 $\rm fb^{-1}$}
\end{figure}

Next the energy spectra for the neutral-current interactions of three flavors of neutrinos 
are shown in Fig.~\ref{fig4} for a number of values for $M_{Z'}$. The expected number of 
NC events of three flavors 
of neutrinos versus $E_\nu > 10 \rm~GeV$ for various $Z'$ mass in FASER$\nu$ can be obtained from 
the corresponding energy spectrum. The highest number of NC events was reported in the $\nu_{\mu}$ 
channel, while the lowest number of NC events in the $\nu_{\tau}$ channel. 
In Fig.~\ref{fig4},  we sum up the contributions from both neutrino and anti-neutrino events. 
Here we have assumed a benchmark detector made of tungsten with dimensions 
$\rm 25\, cm\times 25\, cm \times1\, m$ at the 14 TeV LHC with an integrated luminosity of 
$L = \rm 150 \, fb^{-1}$. 
We use the neutrino fluxes and energy spectra obtained in \cite{FASER:2019dxq} to study the neutrinos 
that pass through FASER$\nu$. We find that muon neutrinos 
are mostly produced from charged-pion decays, electron neutrinos from hyperon, kaon, and $D$-meson 
decays, and tau neutrinos from $D_s$ meson decays. With average energies ranging 
from 600 GeV to 1 TeV, the spectra of the three neutrino flavors cover a broad energy range.

To estimate the sensitivity reach in the parameter space $(g_\nu q_q)$ 
of the $Z'$ model, we first calculate 
the predicted number of events $N_{\rm BSM}$ for the $Z'$ model and the SM number of events 
$N_{\rm SM}$, and treat the statistical error as $\sqrt{N_{\rm BSM}}$
and systematic uncertainty $\sigma_{\rm norm}$ as a fraction ($\sigma_{\rm norm} =20\%, 5\%$)
of the normalization of the SM predictions.
We then define the measure of $\chi^2$ as a function of $(g_\nu q_q)$ and a nuisance 
parameter $\alpha$ as follows \cite{Davidson:2003ha}: 
\begin{eqnarray}\label{Eq.11}
  \chi^2 ( g_q g_\nu, \alpha) &=& \min_{\alpha} \Biggr [ 
   \frac{(N^{\nu_e}_{BSM}-(1+\alpha)N^{\nu_e}_{SM})^2}{N^{\nu_e}_{BSM}} + \frac{(N^{\nu_\mu}_{BSM}-(1+\alpha)N^{\nu_\mu}_{SM})^2}{N^{\nu_\mu}_{BSM}} \nonumber \\
   &&+ \frac{(N^{\nu_\tau}_{BSM}-(1+\alpha)N^{\nu_\tau}_{SM})^2}{N^{\nu_\tau}_{BSM}}
   + \left(\frac{\alpha}{\sigma_{norm}} \right)^2 \Biggr ] \;,
\end{eqnarray}
\begin{figure}[ht!]
	\centering
	\includegraphics[width=17cm,height=8.5cm]{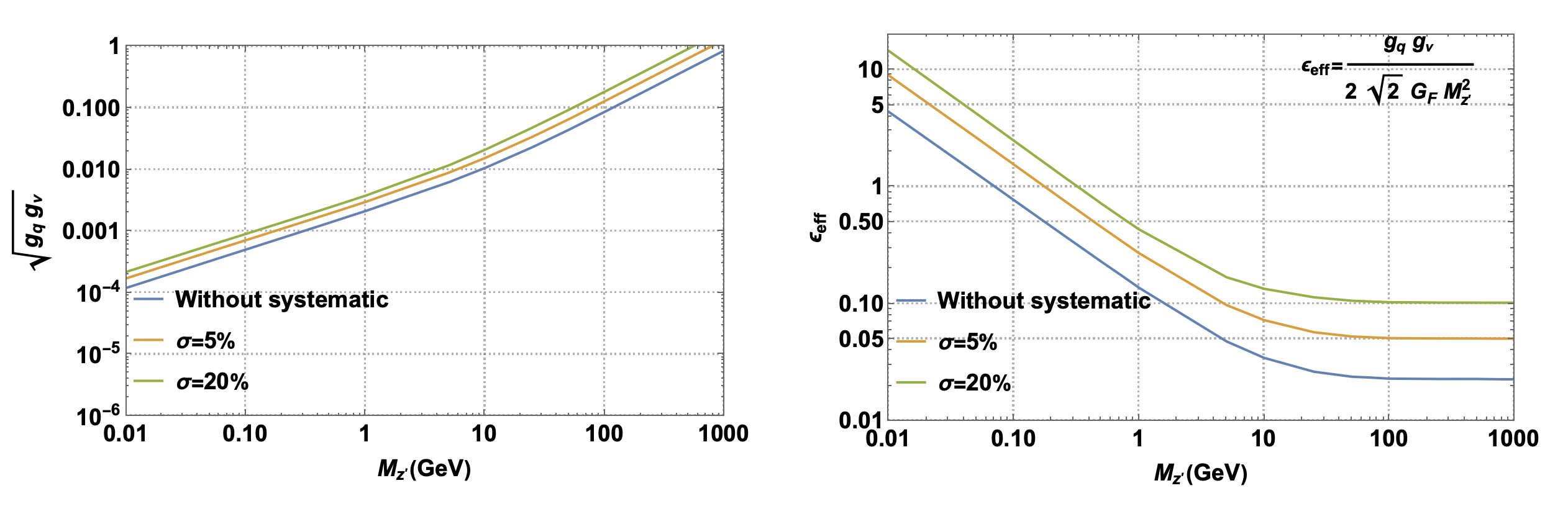}
	\caption{\small \label{fig5}
	Left: Sensitivity reach on the product of couplings $\sqrt{g_q g_\nu}$ versus 
	the $Z'$ mass achieved at FASER$\nu$. 
	Right: Sensitivity reach in terms of $\epsilon_{eff}$ of the NSI's translated from $(g_q g_\nu)$ using  Eq.~(\ref{eps}).
	Systematic uncertainty $\sigma_{\rm norm} = 5,20\%$ and without systematic  uncertainties are shown.
	}
\end{figure}
where $N_{\rm BSM} = N_{Z'} + N_{\rm int} + N_{\rm SM}$ and the minimization is over the
nuisance parameter $\alpha$. Here $N_{Z'}$ is the number of events from the $Z'$ diagram only,
$N_{\rm int}$ is the interference term.
Here we have treated the systematic uncertainties in each neutrino flavor to be the same and use
only one nuisance parameter $\alpha$. Physics-wise the systematic uncertainties come from 
theoretical calculations, the flux of neutrinos from the ATLAS IP, detector response, etc.
We show in Fig.~\ref{fig5} (Left) the 95\% C.L. sensitivity reach (corresponding to $\chi^2 = 3.84$) 
of the product $\sqrt{g_q g_\nu}$ versus $M_{Z'}$ at FASER$\nu$. The higher the systematic uncertainty
the weaker the limit on $\sqrt{g_q g_\nu}$ will be. Nevertheless, the differences among
$\sigma_{\rm norm} = 5\%, 20\%$ and without systematic uncertainties  are relatively small.  The sensitivity reach on
$\sqrt{g_q g_\nu} $ is the best at very small $M_{Z'}$ around $10^{-4}$ at $M_{Z'} = 0.01$ GeV 
and reduces to about $1$  at $M_{Z'} = 1000$ GeV. 
Now we can translate the bounds on $\sqrt{g_q g_\nu}$ to $\epsilon_{eff}$ using Eq.~(\ref{eps}). The bounds on $\epsilon_{eff}$ are shown on the right panel of Fig.~\ref{fig5}. We could see the best limit of $\epsilon_{eff}$ occurs 
at $M_{Z'} \sim 100$ GeV irrespective of the choice of $\sigma_{\rm norm}$.
The curve without systematic uncertainties is giving the best limit of $\epsilon_{eff}$ in the whole $M_{Z'}$ space. 
The limit on $\epsilon_{eff}$ is clearly getting stronger 
as $M_{Z'}$ increases from 0.01 to 100 GeV, but staying 
flat after $M_{Z'}=$100 GeV onward.
The monojet study also shows similar behavior of $\epsilon_{eff}$ at the higher $M_{Z'}$ region.

\begin{figure}[ht!]
	\centering
	\includegraphics[width=16cm,height=10cm]{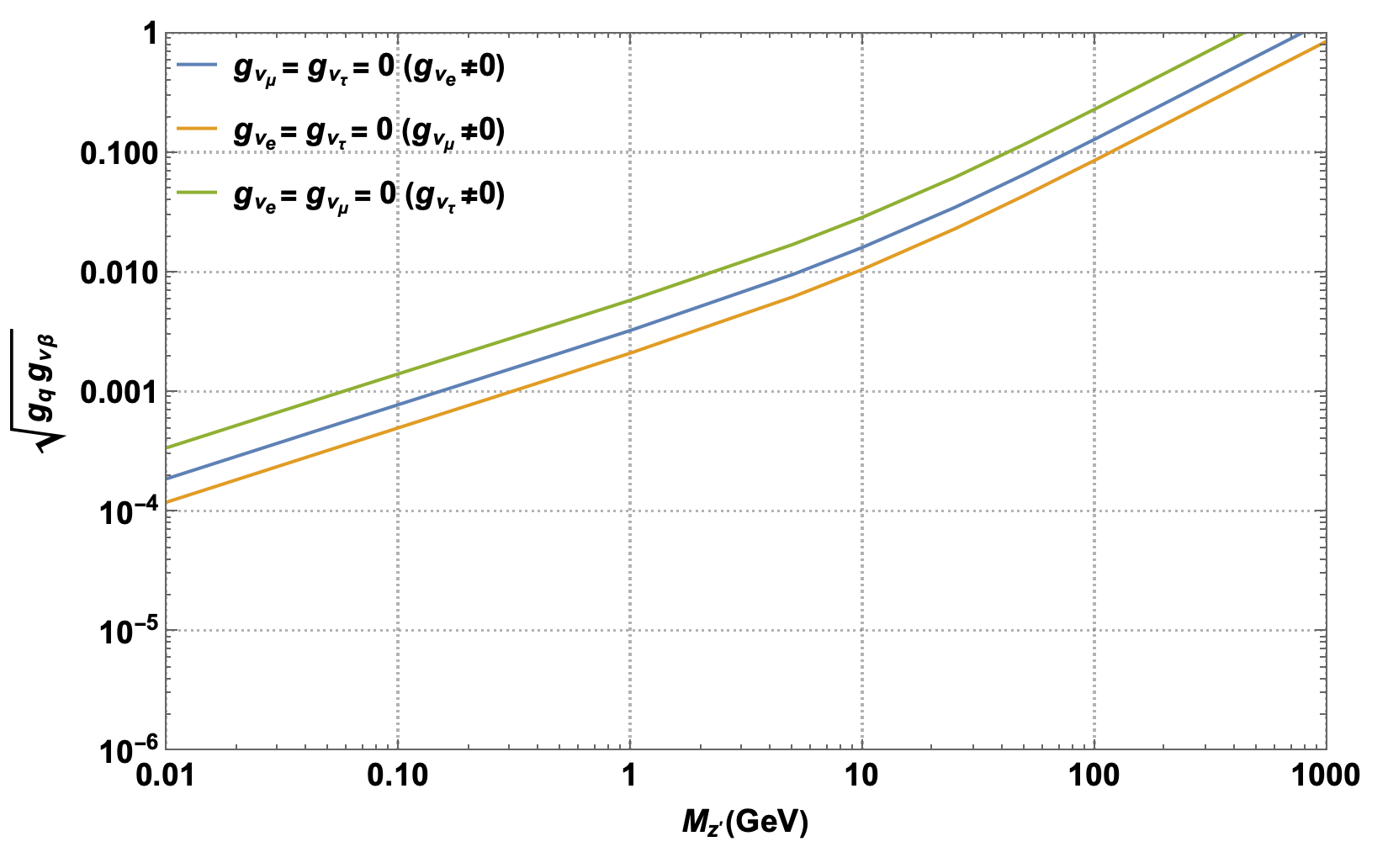}
	\caption{\small \label{fig6}
     Sensitivity reach on the product of couplings $\sqrt{g_q g_{\nu_{\beta}}}$ for
     each neutrino flavor $\beta = e, \mu, \tau$ versus the $Z'$ mass at FASER$\nu$
     (without systematic uncertainties).}
\end{figure}

FASER$\nu$ is primarily designed for the purpose of identifying the flavors of
neutrinos~\cite{Bahraminasr:2020ssz,Bakhti:2020szu}. 
The expected sensitivity for each flavor at FASER$\nu$ is shown in Fig.~\ref{fig6}.
We only show the curves with no systematic uncertainty included. The curves can be compared to
the corresponding one ``Without Systematic" of Fig.~\ref{fig5}.
For the evaluation of $\chi^2$, we consider the special case of Eq.(\ref{Eq.11}) with $\alpha=0$ and it reduces to
\begin{equation}\label{Eq.12}
    \chi^2(g_q g_{\nu_{\beta}}) = [\frac{(N^{\nu_\beta}_{BSM}-N^{\nu_\beta}_{SM})^2}{N^{\nu_\beta}_{BSM}}]
\end{equation}
where $\beta=e,\mu,\tau$.
The green and orange curves of Fig.~\ref{fig6} depict the sensitivity reach of $\sqrt{g_q g_{\nu_{\mu}}}$ and 
$\sqrt{g_q g_{\nu_{e}} }$ ($\chi^2=3.84$) versus $M_{Z'}$, respectively, while the blue curve represent 
the sensitivity reach of $\sqrt{g_q g_{\nu_{\tau}}}$ versus $M_{Z'}$.

It is clear from the Fig.~\ref{fig5} and  Fig.~\ref{fig6} that overall sensitivity reach of $\sqrt{g_q g_\nu}$ 
is dominated by $\sqrt{g_q g_{\nu_{\mu}}}$.
For each neutrino flavor the sensitivity reach on $\sqrt{g_q g_{\nu_\beta}}$ hitting unity at different $Z'$ mass: (i) $\sqrt{g_q g_{\nu_\mu}}$ (orange curve) approaches to 1 at $M_{Z'}\sim$ 1000 GeV, (ii) $\sqrt{g_q g_{\nu_e}}$ (blue curve) approaches to 1 at $M_{Z'}\sim$ 800 GeV, 
and (iii) $\sqrt{g_q g_{\nu_\tau}}$ (green curve) reaches the unity faster than the other two flavors at $M_{Z'}\sim 500$ GeV.

\section{Complementarity of Monojet and FASER$\nu$ Results}

\begin{figure}[H]
	\centering
	\includegraphics[width=17cm,height=10cm]{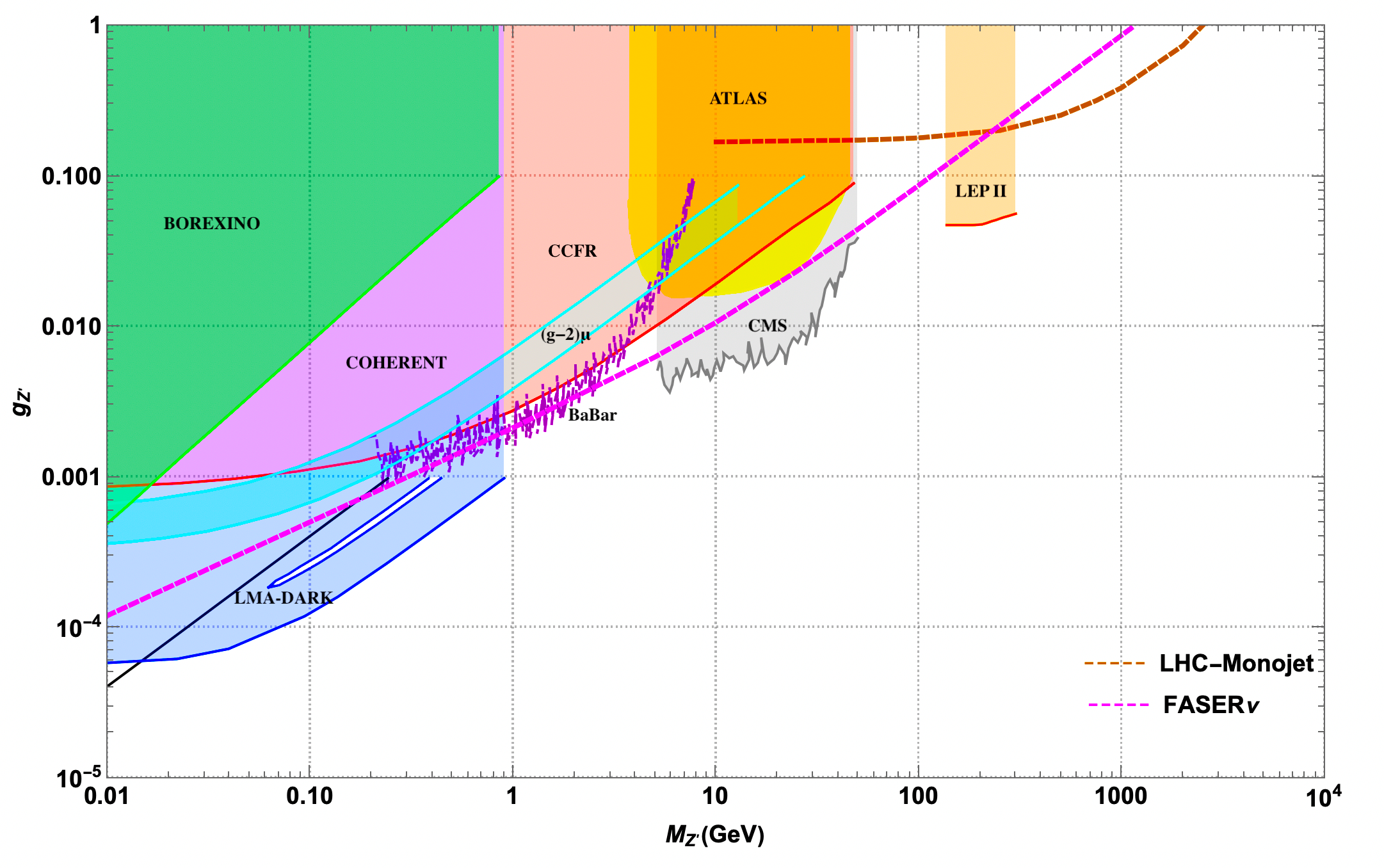}
	\caption{ \small \label{fig7}
	Future sensitivity reach at FASER$\nu$ and the most updated constraint due to monojet production at the LHC at 95\% C.L. 
	Other existing constraints at 95\% C.L. shown include
    (i) the CCFR measurement (red area) of the neutrino trident cross-section~\cite{Ballett:2019xoj},
        (ii) the search of SM $Z$ boson decay to 4 charged leptons in CMS~\cite{Altmannshofer:2014pba}
        (gray area) and ATLAS~\cite{Altmannshofer:2016jzy,ATLAS:2014jlg} (yellow area) reinterpreted under the hypothesis 
        of $Z \rightarrow Z'\mu\mu$,
    (iii) the search of $e^+ e^- \rightarrow \mu^+\mu^-Z',\; Z'\rightarrow  \mu ^+\mu^-$ from BaBar~\cite{CMS:2018yxg} (purple area),
    (iv) bounds from Borexino~\cite{BaBar:2016sci,Kamada:2015era}  (green area), 
    (v) $(g-2)_\mu~2\sigma$ band related to the anomalous magnetic moment of 
        muon~\cite{Gninenko:2020xys} (cyan area),
    (vi) the constraint from the present COHERENT data~\cite{Cadeddu:2020nbr,Denton:2018xmq,Liao:2017uzy} (blue area),  
        (vii) the LMA-DARK solution~\cite{Cadeddu:2020nbr} with $x=0$ (with $x=2$)
        (black curve), and (viii) the constrain from LEP II \cite{Buckley:2011vc,Carena:2004xs} (orange area). 
	}
\end{figure}

Monojet production at the LHC and the NC deep-inelastic scattering at FASER$\nu$ 
cover different energy scales. It would be useful to put both results together.  
We show in Fig.~\ref{fig7} the future sensitivity reach at FASER$\nu$ and the most
updated constraints due to monojet production at the LHC. It is interesting to see
that FASER$\nu$ is mostly sensitive to small $M_{Z'}$ region from $10^{-2} - O(100)$ GeV
while monojet production is more sensitive for $M_{Z'} \agt 100$ GeV to a few TeV.
In Fig.~\ref{fig7}, we also include other existing constraints at 95\% C.L., including
(i) the CCFR measurement of the neutrino trident cross-section~\cite{Ballett:2019xoj},
(ii) the search of SM $Z$ boson decay to 4 charged leptons in CMS~\cite{Altmannshofer:2014pba}
and 
   ATLAS~\cite{Altmannshofer:2016jzy,ATLAS:2014jlg} reinterpreted under the hypothesis of $Z \rightarrow Z'\mu\mu$,
(iii) the search of $e^+ e^- \rightarrow \mu^+\mu^-Z',$ followed by  
$Z'\rightarrow  \mu ^+\mu^-$ from BaBar\cite{CMS:2018yxg},
(iv) bounds from Borexino~\cite{BaBar:2016sci,Kamada:2015era},
   (v) $(g-2)_\mu~2\sigma$ band related to the anomalous magnetic moment of
   muon~\cite{Gninenko:2020xys},
(vi) the constraint from the present  COHERENT data~\cite{Cadeddu:2020nbr,Denton:2018xmq,Liao:2017uzy}, 
(vii) the LMA-DARK solution~\cite{Cadeddu:2020nbr} with $x=0$ (with $x=2$),
and
(viii) the LEP II bounds on couplings to electrons derived from
\cite{Buckley:2011vc,Carena:2004xs}, where
we have assumed a single fermion helicity in the $Z'$ coupling. 
The constraints on the couplings of the $Z'$ to leptons are significantly more stringent than those to quarks. 
In particular, the process $e^+ e^-\rightarrow Z' \rightarrow e^+ e^-$ leads to a constraint of $g_{ee}^{Z'} \leq 0.044\times (M_{Z'}/\rm 200 GeV)$ 
for $Z'$ masses above roughly 200 GeV.

In the intermediate mass range ($1\,{\rm GeV} \alt M_{Z'} < 50\, {\rm GeV} $),
the FASER$\nu$'s sensitivities are comparable with the existing constraints,
except for the range $M_{Z'} = 5 -50$ GeV, where the CMS and ATLAS searches on
SM $Z$ boson decay into 4 charged leptons are somewhat better. In the low mass regime ($M_{Z'}=0.01-1 \rm GeV)$ the COHERENT results are better than the FASER$\nu$ sensitivites.
The LMA-DARK solution is also better than FASER$\nu$ sensitivites in $0.01\, \rm GeV< M_{Z'}\leq 0.1\, \rm GeV$ region, however in the higher $M_{Z'}$ region ($M_{Z'}>0.1\, \rm GeV$) FASER$\nu$ can constrain better than the LMA-DARK.
In the high mass regime (100 GeV $\alt M_{Z'}$),  the LHC Monojet results
constrain better than the sensitivities offered by the FASER$\nu$, wheree we can see
the crossover between FASER$\nu$ and LHC-monojet results at $M_{Z'}\sim$250 GeV. 

Here we make a brief comparison with the sensitivity achieved at the DUNE near-detector.
The $\nu-e$ scattering sensitivity to the $L_e-L_\mu~Z'$ model at 90\% C.L. 
was performed in Ref.~\cite{He:1991qd}, and 
the dimuon neutrino trident sensitivity to the $L_\mu-L_\tau$ model~\cite{He:1991qd}
with no kinetic mixing at 90\% C.L. were reported in Ref.~\cite{Ballett:2019xoj}. 
Sensitivity on $g'$ with $L_e-L_\mu~Z'$ model reaches the best at very small $M_{Z'}$ around $\sim 5\times10^{-5}$ at $M_{Z'}$=0.01 GeV and rises 
to about 0.01 when $M_{Z'}$=10 GeV. 
For the case of  $L_\mu-L_\tau$ model the $g'$ value with $M_{Z'}$=0.01 is 
in the order of $\sim 2\times10^{-4}$ and rises to $\sim$ 0.01 at $M_{Z'}$=10 GeV. 
On the other hand, the best FASER$\nu$ sensitivity that we can achieve is
$\sqrt{g_q g_\nu} \sim 10^{-4}$ at $M_{Z'} = 0.01$ GeV 
and rises to about $0.01$  at $M_{Z'} = 10$ GeV.
Therefore, we can see that the FASER$\nu$ sensitivity is comparable to that of
DUNE.

\section{Conclusions}

In this paper, we have studied the neutral-current scattering between
neutrinos and nuclei in the FASER$\nu$ detector and calculated the 
expected sensitivity reach on possible NSI's using a simplified $Z'$ model.
We investigated the advantage of FASER$\nu$ in wide mass range search for $Z'$ and to determine the flavor dependence of the coupling between neutrino and this new boson, for which we found that FASER$\nu$ is sensitive to $g_{\nu_\mu}$ because of the larger statistics. We also found that the impact of systematical uncertainty due to normalization is relatively small in the smaller $M_{Z'}$ region.

We have also investigated the effects of the simplified $Z'$ model on
monojet production at the LHC, followed by an update on the existing
bound using the most recent results on monojet production at the LHC
with 139 fb$^{-1}$ luminosity. We have found substantial improvement over
previous works.

While the FASER$\nu$ can achieve the best sensitivity at small $M_{Z'}$ regime, the sensitivity using monojet production,  on the other hand, is more profound at high mass region. Thus, complementarity in mass range
coverage is established.
Overall, the FASER$\nu$ offers a sensitivity reach better 
than the existing constraints at low mass region ($M_{Z'} < 0.1$) GeV, except for the COHERENT constraint and for the DUNE near-detector 
$\nu-e$ scattering sensitivity.
The FASER$\nu$ sensitivity is comparable to existing constraints 
in the intermediate mass region ($0.1 \alt M_{Z'} \alt 10$ GeV).
We explored the capability of thee FASER$\nu$ detector to discern
individual neutrino flavors. 
In both the FASER$\nu$ experiment and LHC we obtained the best limit for $\epsilon_{eff}$ based on the translation of $\sqrt{g_q g_\nu}$.
Further full detector simulation
at FASER$\nu$ is called for establishing the feasibilty.

More and more particle-physics experiments or cosmological observatories provide the bounds at the lower mass region of $Z'$. However, we have not seen any signal so far. One may be more interested in the heavier $Z'$ models, for which FASER$\nu$ and monojet play an
important role in that search. We are looking forward to the upgrade of FASER/FASER$\nu$, which is being discussed in the collaboration group. 
For a complete picture of $Z'$ search, our suggestion is to cover 
the mass range around $100$ GeV and those above $\sim 300$ GeV, 
which are still lack of constraints.

\section*{Acknowledgement}
Special thanks to Felix Kling, Shih-Chieh Hsu and Zeren Simon Wang for enlightening discussion. Also thanks to Olivier Mattelaer for a wonderful usage of Madgraph. TC acknowledges the support from National Center for Theoretical Sciences. The work was supported in part by Taiwan MoST with grant no. MOST-110-2112-M-007-017-MY3.

\end{document}